\documentclass[runningheads,openany]{svmult}

\usepackage{makeidx}   
\usepackage{graphicx}  
\usepackage{subeqnar}  
\usepackage{multicol}  
\usepackage{physmubb}  
\usepackage{rotating,epsfig} 
\usepackage{bm}        
\usepackage{url}       
\usepackage{floatflt}
\usepackage{amsmath}
\usepackage{amssymb}
\usepackage{slashed}

\usepackage{harvard}
\harvardparenthesis{square}

\usepackage{booktabs}  
\usepackage{dcolumn}
\newcolumntype{d}{D{.}{.}{-1}}
\makeindex             

\include{macros}

\begin{document}

\frontmatter



\mainmatter

\title{On the Combination of TDDFT with Molecular Dynamics: New Developments}
\titlerunning{TDDFT and MD: New Developments}
\author{Jos\'e L. Alonso, Alberto Castro, Pablo Echenique and Angel Rubio}
\authorrunning{J.\protect\,L. Alonso, A. Castro, P. Echenique and A. Rubio}
\tocauthor{J.\protect\,L. Alonso, A. Castro, P. Echenique and A. Rubio}

\maketitle
\label{chap_1.1}

\textbf{Jos\'e L. Alonso} \\
Departamento de F{\'{\i}}sica Te{\'{o}}rica, Universidad de Zaragoza, Pedro Cerbuna 12, E-50009 Zaragoza (Spain);
Instituto de Biocomputaci\'on y F\'{\i}sica de Sistemas Complejos (BIFI), Universidad of Zaragoza, Mariano Esquillor s/n, E-50018 Zaragoza (Spain). \\
\texttt{alonso.buj@gmail.com}

\textbf{Alberto Castro} \\
Instituto de Biocomputaci\'on y F\'{\i}sica de Sistemas Complejos (BIFI), Universidad of Zaragoza, Mariano Esquillor s/n, E-50018 Zaragoza (Spain). \\
\texttt{acastro@bifi.es}

\textbf{Pablo Echenique} \\
Instituto de Qu{\'{\i}}mica F{\'{\i}}sica ``Rocasolano'', CSIC, Serrano 119, E-28006 Madrid (Spain); 
Instituto de Biocomputaci\'on y F\'{\i}sica de Sistemas Complejos (BIFI), Universidad of Zaragoza, Mariano Esquillor s/n, E-50018 Zaragoza (Spain); 
Departamento de F{\'{\i}}sica Te{\'{o}}rica, Universidad de Zaragoza, Pedro Cerbuna 12, E-50009 Zaragoza (Spain). \\
\texttt{echenique.p@gmail.com}

\textbf{Angel Rubio} \\
Nano-Bio Spectroscopy Group and ETSF Scientific Development Centre, Departamento de F{\'{\i}}sica de Materiales, Universidad
del Pa{\'{\i}}s Vasco, E-20018 San Sebasti{\'{a}}n (Spain); 
Centro de F{\'{\i}}sica de Materiales CSIC-UPV/EHU-MPC and DIPC, E-20018 San Sebasti{\'{a}}n (Spain); 
Fritz-Haber-Institut der Max-Planck-Gesellschaft, Faradayweg 4-6 , D-14195 Berlin-Dahlem, Germany \\
\texttt{angel.rubio@ehu.es}

\hspace{1cm}


\section{Introduction}


In principle, we should not need the time-dependent extension of
density-functional theory (TDDFT) for excitations, and in particular
not for Molecular Dynamics (MD) studies: the theorem by Hohenberg and
Kohn~\cite{hohenberg1964} teaches us that for any observable that we
wish to look at (including dynamical properties or observables
dependent on excited states) there is a corresponding functional of
the ground-state density. Yet the unavailability of such magic
functionals in many cases (the theorem is a non-constructive existence
result) demands the development and use of the alternative exact
reformulation of quantum mechanics provided by TDDFT. This theory
defines a convenient route to electronic excitations and to the
dynamics of a many-electron system subject to an arbitrary
time-dependent perturbation (discussed in previous chaters of this
book).  This is, in fact, the main purpose of inscribing TDDFT in a MD
framework --the inclusion of the effect of electronic
excited states in the dynamics. However, as we will show in this chapter, it may not
be the only use of TDDFT in this context.

The term ``Ab Initio Molecular Dynamics'' (AIMD) has been exclusively
identified in the past with the Car-Parrinello (CP)
technique~\cite{car1985}. This method combines ground-state DFT with
MD, providing an efficient reformulation of ground-state
Born-Oppenheimer MD (gsBOMD)~\cite{marx2000}. However, the ``AIMD''
words have broader meaning, and should include all the possible MD
techniques that make use of a first principles approach to tackle the
many-electron problem. For example, Ehrenfest MD (EMD) can also be one
AIMD scheme if TDDFT is used to propagate the electronic
subsystem. This is the most common manner in which TDDFT and MD have
been combined in the past: as a means to study fast out-of-equilibrium
processes, typically intense laser irradiations or ionic
collisions~\cite{saalmann1996,saalmann1998,reinhard1999,kunert2001,castro2004}.

Nevertheless, there are other possibilities. In this chapter we review two
recent proposals: In Section~\ref{section:fast-ehrenfest-dynamics}, we show
how TDDFT can be used to design efficient gsBOMD
algorithms~\cite{alonso2008,andrade2009} --even if the electronic excited
states are in this case not relevant. The work described in
Section~\ref{section:md-finite-temperature} addresses the problem of mixed
quantum-classical systems at thermal equilibrium~\cite{alonso2010}.

\section{Fast Ehrenfest Molecular Dynamics}
\label{section:fast-ehrenfest-dynamics}

Ehrenfest Molecular Dynamics (EMD) is a model for describing the evolution of
a mixed quantum-classical system. The equations of motion are given by:
\begin{eqnarray}
\label{eq:ehrenfest-1}
M_\alpha \frac{{\rm d}^2}{{\rm d}t^2}\vec{R}_\alpha(t) & = & - 
\langle\Phi(t)\vert \nabla_{\vec{R}_\alpha}\hat{H}_{\rm e}(\vec{R}(t),t)
\vert\Phi(t)\rangle\,,
\\
\label{eq:ehrenfest-2}
\I \frac{{\rm d}}{{\rm d}t}\vert\Phi(t)\rangle & = & 
\hat{H}(\vec{R}(t),t)\vert\Phi(t)\rangle\,,
\end{eqnarray}
where $\Phi(t)$ is the state of the quantum subsystem (we will assume that this
is a set of $N$ electrons), and $\lbrace \vec{R}_\alpha
\rbrace_{\alpha=1}^{N_{\rm nuc}}$ are the position coordinates of $N_{\rm
nuc}$ classical particles (a set of $N_{\rm nuc}$ nuclei of masses $M_\alpha$
and charges $z_\alpha$). The quantum (or \emph{electronic}) Hamiltonian
operator $\hat{H}_{\rm e}(\vec{R},t)$ depends on these classical coordinates,
and is usually given by:
\begin{eqnarray}
\label{eq:He}
\nonumber
\hat{H}_{\rm e}(\vec{R},t) & := & \sum_{i=1}^N \frac{1}{2}\hat{p}_i^2  + 
\sum_{i,j<i}\frac{1}{\vert\hat{\vec{r}}_i - \hat{\vec{r}}_j\vert}
+ \sum_{\beta<\alpha}\frac{z_\alpha z_\beta}{\vert{\vec{R}}_\alpha - 
{\vec{R}}_\beta \vert}
\\
& & 
- \sum_{\alpha,i}\frac{z_\alpha }{\vert{\vec{R}}_\alpha -
 \hat{\vec{r}}_i \vert}
+ \sum_{i} v^{\rm e}_{\rm ext}(\hat{\vec{r}}_i,t) 
+ \sum_{\alpha} v^{\rm n}_{\rm ext}(\vec{R}_\alpha,t)\,,
\end{eqnarray}
where $v^{\rm e}_{\rm ext}$ and $v^{\rm n}_{\rm ext}$ are external potentials
acting on the electrons and nuclei, respectively \cite{echenique2007}. Atomic
units are used throughout the document in order to get rid of constant factors
such as $\hbar$ or $1/4\pi\varepsilon_0$.

Given this definition, one can show that Eq.~(\ref{eq:ehrenfest-1}) can be
rewritten as:
\begin{equation}
\label{eq:ehrenfest-density}
M_\alpha \frac{{\rm d}^2}{{\rm d}t^2}\vec{R}_\alpha(t) =
 - \int\!\!{\rm d}^3r\; n(\vec{r},t)
\nabla_{\vec{R}_\alpha} v_0(\vec{r},\vec{R}(t))\,,
\end{equation}
where
\begin{eqnarray}
\nonumber
v_0(\vec{r},\vec{R}) & := &
-\sum_{\alpha}\frac{z_\alpha }{\vert{\vec{R}}_\alpha - \vec{r} \vert}
\\
& &
+ \frac{1}{N}\sum_{\alpha} v^{\rm n}_{\rm ext}(\vec{R}_\alpha,t)
+ \frac{1}{N}\sum_{\beta<\alpha}\frac{z_\alpha z_\beta}{\vert{\vec{R}}_\alpha 
 - {\vec{R}}_\beta \vert} \,;
\end{eqnarray}
a result which is known as the ``electrostatic force theorem'' in the
quantum chemistry literature \cite{levine2000}, and which is based in
the fact that the gradient $\nabla_{\vec{R}_\alpha} \hat{H}_{\rm
  e}(\vec{R}(t),t)$ is a one-body local multiplicative operator (as
far as the electrons are concerned), i.e., it is a sum of one-electron
operators whose action amounts to a multiplication in real
space~\cite{eschrig2003,vonbarth2004}.

Eq.~(\ref{eq:ehrenfest-density}) shows that the knowledge of the
time-dependent electronic density $n(\vec{r},t)$ suffices to obtain the
nuclear movement. This fact is the basis for TDDFT-based Ehrenfest MD
(E-TDDFT): instead of solving Eq.~(\ref{eq:ehrenfest-2}), we solve the
corresponding time-dependent Kohn-Sham system, which provides an approximation
to $n(\vec{r},t)$:
\begin{equation}
\label{eq:tdks}
\I\frac{\partial}{\partial t}\varphi_j(\vec{r},t) =  
-\frac{1}{2}\nabla^2\varphi_j(\vec{r},t) 
+ v_{\rm KS}[n](\vec{r},t)\varphi_j(\vec{r},t)\,,
\quad j=1,\ldots,N\,,
\end{equation}
being
\begin{equation}
\label{eq:vks}
v_{\rm KS}[n](\vec{r},t) := 
\sum_{\alpha}\frac{-z_\alpha }{\vert{\vec{R}}_\alpha(t) - \vec{r} \vert}
+ v_{\rm H}[n](\vec{r},t) 
+ v_{\rm xc}[n](\vec{r},t)  + v^{\rm e}_{\rm ext}(\vec{r},t)\,,
\end{equation}
and
\begin{equation}
\label{eq:density}
n(\vec{r},t) := 2\sum_{j=1}^N \vert\varphi_j(\vec{r},t)\vert^2\,,
\end{equation}
where $v_{\rm KS}[n](\vec{r},t)$ is the time-dependent Kohn-Sham potential,
and $v_{\rm H}[n](\vec{r},t)$ and $v_{\rm xc}[n](\vec{r},t)$ are the Hartree
and exchange-correlation potential, respectively. For simplicity, we assume an
even number of electrons in a spin-compensated configuration.

The equations of motion for E-TDDFT (\ref{eq:ehrenfest-density}, \ref{eq:tdks}
and \ref{eq:density}) can be derived from the following Lagrangian (assuming an adiabatic
approximation for the exchange and correlation potential, as it is commonly done
in practical implementations of TDDFT):
\begin{eqnarray}
\nonumber
L_\mu[\varphi,\dot{\varphi},\vec{R},\dot{\vec{R}}] & := & 
 \mu\frac{\I}{2}\sum_j \left(\langle \varphi_j \vert \dot{\varphi_j}\rangle
- \langle \dot{\varphi}_j \vert \varphi_j\rangle\right)
\\\label{eq:lagrangian-etddft}
& & 
+ \sum_\alpha \frac{1}{2}M_\alpha \dot{\vec{R}}^2_\alpha -
  E_{\rm KS}[\varphi,\vec{R}]\,,
\end{eqnarray}
for $\mu=1$ (the reason for including this parameter $\mu$ will become clear
in what follows). We use a dot to denote time-derivatives.

The term $E_{\rm KS}$ is the Kohn-Sham ground-state energy functional:
\begin{align}
\nonumber
E_{\rm KS}[\varphi,\vec{R}] := 2\sum_j
\langle \varphi_j\vert\frac{\hat{p}^2}{2}\vert\varphi_j\rangle
- \int\!\!{\rm d}^3r\;\sum_{\alpha}\frac{z_\alpha }{\vert{\vec{R}}_\alpha - 
 \vec{r} \vert}n(\vec{r})
\\
+ \frac{1}{2}\int\!\!{\rm d}^3r\; v_{\rm Hartree}[n](\vec{r})n(\vec{r}) + 
 E_{\rm xc}[n] + 
\sum_{\beta<\alpha}\frac{z_\alpha z_\beta}{\vert{\vec{R}}_\alpha - 
{\vec{R}}_\beta \vert}\,.
\end{align}

Note that, when the time-dependent orbitals are introduced into this
expression (as it is done in E-TDDFT), it becomes a functional of the
Kohn-Sham orbitals at each time, and not a functional of the ground state
density. Also, from here on, we assume that there are no external potentials
$v_{\rm ext}^{\rm e}$ and $v_{\rm ext}^{\rm n}$, since they do not add
anything to the following discussion.

It is worth remarking now that EMD differs from gsBOMD, and it is instructive
to see in which way. We do so in the initial formulation of
Eqs.~(\ref{eq:ehrenfest-1}) and~(\ref{eq:ehrenfest-2}) using the $N$-electron
wavefunction for simplicity, i.e., we forget for a moment the TDDFT formalism.

To illustrate the main concepts we start by projecting the EMD equations into the adiabatic basis, formed at
each nuclear configuration by the set of eigenfunctions of the electronic
Hamiltonian:
\begin{equation}
\label{eq:adiabatic-basis}
\hat{H}_{\rm e}(\vec{R})\vert \Psi_m(\vec{R})\rangle = 
E_m(\vec{R})\vert\Psi_m(\vec{R})\rangle\,,
\end{equation}
\begin{equation}
\vert\Phi(t)\rangle = \sum_m c_m(t)\vert\Psi_m(\vec{R}(t))\rangle\,.
\end{equation}
The result is:
\begin{eqnarray}
\nonumber
M_\alpha \frac{{\rm d}^2}{{\rm d}t^2}\vec{R}_\alpha(t) & = & 
-\sum_m \vert c_m(t)\vert^2 \nabla_{\vec{R}_\alpha} E_m(\vec{R}(t))
\\
& & 
- \sum_{mn} c^*_m(t)c_n(t) \left[ E_m(\vec{R}(t)) - E_n(\vec{R}(t)) \right] \vec{d}_\alpha^{mn}(\vec{R}(t))\,.
\\
\I \frac{{\rm d}}{{\rm d}t} c_m(t) & = & E_m(\vec{R}(t)) c_m(t) 
-\I \sum_n c_n(t) \left[
\sum_\alpha \dot{\vec{R}}_\alpha \cdot \vec{d}_\alpha^{mn}(\vec{R}(t))
\right]
\end{eqnarray}
where the ``non-adiabatic couplings'' are defined as:
\begin{equation}
\vec{d}_\alpha^{mn}(\vec{R}) :=  \langle \Psi_m(\vec{R})
 \vert \nabla_{\vec{R}_\alpha} \Psi_n(\vec{R}) \rangle\,.
\end{equation}

If these are negligible, and we assume that the electronic system starts
from the ground state ($c_m(0)=\delta_{m0}$), EMD reduces to gsBOMD:
\begin{eqnarray}
\label{eq:gsbomd-1}
M_\alpha \frac{{\rm d}^2}{{\rm d}t^2}\vec{R}_\alpha(t) & = & 
\nabla_{\vec{R}_\alpha} E_0(\vec{R}(t))\,,
\\
\label{eq:gsbomd-2}
c_m(t) & = & \delta_{m0}\,.
\end{eqnarray}

Now, in order to integrate the gsBOMD equations, one can make use of
ground-state DFT, since the only necessary ingredient is the ground-state
energy $E_0(\vec{R}(t))$. One could thus precompute this hyper-surface, in
order to propagate the nuclear dynamics \emph{a posteriori}, or else only
compute the energies at the $\vec{R}$ points visited by the dynamics (a
procedure normally known as ``on-the-fly''). However, Car and
Parrinello~\cite{car1985} proposed an alternative, based on the following
Lagrangian:
\begin{eqnarray}
\nonumber
L^{\rm CP}_{\lambda}[\varphi,\dot{\varphi},\vec{R},\dot{\vec{R}}] & := & 
\lambda\frac{\I}{2}\sum_j 
	\langle \dot{\varphi}_j \vert \dot{\varphi}_j \rangle 
+ \sum_\alpha \frac{1}{2}M_\alpha \dot{\vec{R}}^2_\alpha
\\
& & 
- E_{\rm KS}[\varphi,\vec{R}]
+\sum_{ij}\Lambda_{ij}\left( \langle \varphi_i \vert \varphi_j \rangle - \delta_{ij}\right)\,.
\end{eqnarray}

Note the presence of a fictitious mass $\lambda$, and of a set of Lagrange
multipliers $\Lambda_{ij}$ associated to the constraints that keep the KS
orbitals orthonormal along the evolution. The Car-Parrinello (CP) equations
that stem from this Lagrangian are:
\begin{equation}
M_\alpha \frac{{\rm d}^2}{{\rm d}t^2}\vec{R}_\alpha(t)  = 
	- \nabla_{\vec{R}_\alpha} E_{\rm KS}[\varphi(t),\vec{R}(t))]\,, 
\end{equation}
\begin{equation}
\lambda \ddot{\varphi}_j(\vec{r},t) = - \frac{1}{2}\nabla^2\varphi_j(\vec{r},t) + v_{\rm KS}[n](\vec{r},t)\varphi_j(\vec{r},t)
+ \sum_k \Lambda_{jk}\varphi_k(\vec{r},t)\,,
\end{equation}
\begin{equation}
\label{eq:orthonormality}
\langle \varphi_i(t) \vert \varphi_j(t) \rangle = \delta_{ij}\,.
\end{equation}

The first of these three sets of equation ensures that CP molecular dynamics
(CPMD) is (approximately) equivalent to gsBOMD if the KS orbitals stay close
to the ground-state ones; the second equation is an auxiliary,
\emph{fictitious} electronic propagation that enforces this proximity to the
ground state for a certain range of values of the ``mass'' $\lambda$; whereas
the last equation demands the constant orthonormality of the electronic
orbitals. Another role of the fictitious mass $lambda$ is to accelerate the
fake electronic dynamics, and as a consequence to improve the numerical
efficiency. This efficiency (in addition to the success of DFT in the
calculation of total energies with chemical accuracy) has made of CPMD the method of choice for
performing ab initio gsBOMD during the last decades.

When attempting simulations of very large systems, the calculations must be done
using the massive parallel architectures presently available, therefore
one must ensure a good scalability of the computational
algorithms with respect to the number of processors and the size of the systems (ie. number of atoms).
The CPMD technique at a given point has to face the problem posed by the need of orthonormalization, as required by
Eq.~(\ref{eq:orthonormality}). This is a very non-local process (regardless of
the algorithm used), and therefore very difficult to parallelize efficiently.
Linear-scaling methods and other approaches have been proposed
recently~\cite{kuhne2007} to improve the speed of the CP technique.

One possibility to circumvent the orthonormalization issue is to do
E-TDDFT (which automatically conserves the orthonormality) instead of
CPMD, for those cases in which the coupling to higher electronic
excited states is weak, and therefore E-TDDFT is almost equivalent to
gsBOMD. This fact was first realized by
Theilhaber~\cite{theilhaber1992}.  Unfortunately, the required time
step for E-TDDFT is very small (two to three orders of magnitude
smaller than the CPMD time-step), which makes it very inefficient
computationally. The reason is that the simulation must follow the
real electronic motion, which is very fast (in contrast to the
fictitious electronic motion used in CPMD). In \cite{alonso2008} and
\cite{andrade2009}, however, it was shown how the time-step can be
increased by modifying the $\mu$ parameter in the definition of the
Lagrangian function given in Eq.~(\ref{eq:lagrangian-etddft}), which
for normal Ehrenfest dynamics should be $\mu = 1$.

For any $\mu$, the equations of motion derived from this Lagrangian function
are:
\begin{eqnarray}
\label{eq:new-ehrenfest-1}
\I\mu\frac{\partial}{\partial t}\varphi_j(\vec{r},t) & = & -\frac{1}{2}\nabla^2\varphi_j(\vec{r},t) 
+ v_{\rm KS}[n](\vec{r},t)\varphi_j(\vec{r},t)\,,
\\
\label{eq:new-ehrenfest-2}
M_\alpha \frac{{\rm d}^2}{{\rm d}t^2}\vec{R}_\alpha(t)  & = & 
- \nabla_{\vec{R}_\alpha} E_{\rm KS}[\varphi(t), \vec{R}(t)]\,.
\end{eqnarray}

The only difference with respect to E-TDDFT is the appearance of the $\mu$
parameter multiplying the time-derivative of the time-dependent KS
equations. The most relevant features of this dynamics are:

\begin{enumerate}

\item The orthogonality of the time-dependent KS orbitals is automatically
 preserved along the evolution, so that there is no need to perform any orthonormalization
procedure.

\item The ``exact'' total energy of the system, defined as
\begin{equation}
\label{eq:total-energy}
E = \frac{1}{2}\sum_\alpha M_\alpha \dot{\vec{R}}^2_\alpha +
  E_{\rm KS}[\varphi,\vec{R}]\,,
\end{equation}
is also preserved along the evolution. Note that it is independent of $\mu$
and it coincides with the same exact energy that is preserved along the gsBOMD
evolution. In contrast, the preserved energy in CPMD is given by:
\begin{equation}
E_{\rm CP} = E + \frac{1}{2}\lambda \sum_j 
 \langle \dot{\varphi}_j \vert \dot{\varphi}_j \rangle\,,
\end{equation}
where $E$, given by Eq.~(\ref{eq:total-energy}), is now time-dependent.
It can be seen how the new constant of motion $E_{\rm CP}$ actually depends on
$\lambda$, which is the fictitious electronic mass introduced in the CP
formulation.

\item If we consider Eq.~(\ref{eq:new-ehrenfest-1}), and write the left hand
side as:
\begin{equation}
\I\mu\frac{\partial}{\partial t} = \I \frac{\partial}{\partial t_\mu}\,,
\end{equation}
the resulting equations can be seen as the standard Ehrenfest method in terms
of a fictions time $t_\mu$. This has the effect of scaling the TDDFT
excitation energies by a $1/\mu$ factor. So we may open or close the
electronic gap by using a $\mu$ smaller or larger than 1. Obviously, if $\mu
\to 0$, then the gap becomes infinite, and we retrieve the adiabatic
(gsBOMD) regime. 

\item The second important effect of this time re-scaling is a change in the
required time-step for the numerical propagation; if the time-step for the
standard E-TDDFT equations is $\Delta t$, then the required time-step for the
new dynamics is $\Delta t_\mu = \mu\Delta t$. In other words, the propagation
will be $\mu$ times faster.

\item Taking into account the two previous points and recalling that the
purpose of this modified Ehrenfest dynamics is to reproduce, albeit
approximately, the gsBOMD results, it becomes clear that there is a tradeoff
affecting the optimal choice for the value of $\mu$: low values (but still
larger than one) will give physical accuracy, while large values will produce
a faster propagation. The optimal value is the maximum value that still keeps
the system near the adiabatic regime. It is reasonable to expect that this
value will be given by the ratio between the electronic gap and the highest
vibrational frequency of the nuclei. For many systems, like some molecules or
insulators, this ratio is large and we can expect large improvements with
respect to standard Ehrenfest MD. For other systems, like metals, this ratio
is small or zero and the new method will not work.

\item Regarding the scaling with the system size, the modified Ehrenfest
dynamics evidently inherits the main advantage of the original one: since
propagation preserves the orthonormality of the KS orbitals, it needs not be
imposed and the numerical cost is proportional to $N_W N_C$ (with $N_W$ the
number of orbitals and $N_C$ the number of grid points or basis set
coefficients). For CPMD, a reorthogonalization has to be done each time step,
so the cost is proportional to $N_W^2 N_C$. From these scaling properties, we
can predict that for large enough systems the Ehrenfest method will be less
costly than CP. For smaller systems, however, this gain will not compensate
for the fact that the time-step, despite being increased by the $\mu$ factor,
will still need to be one or two orders of magnitude smaller then the
time-step utilized in CPMD.

\end{enumerate}

\begin{figure}[t]
\center{\includegraphics[width=0.9\columnwidth]{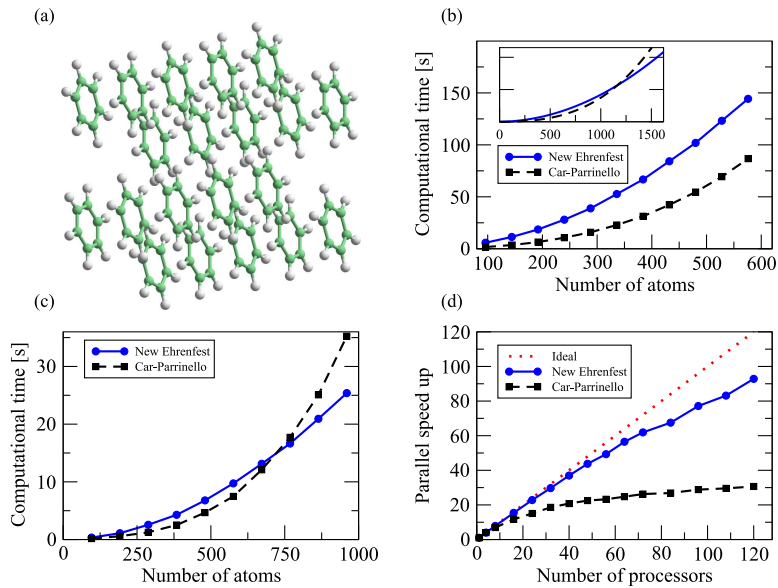}}
\caption{
\label{fig:benzene-crystal}
(a) Scheme of the benzene molecule array.  (b) Single processor
computational cost for different system sizes. (inset) Polynomial
extrapolation for larger systems.  (c) Parallel computational cost for
different system sizes. (d) Parallel scaling with respect to the
number of processor for a system of 480 atoms. In both cases, a mixed
states-domain parallelization is used to maximize the performance.  }
\end{figure}

Some numerical examples, performed with the octopus
code~\cite{marques2003,castro2006}, that give an idea of the
performance of this modified Ehrenfest dynamics were shown in
\cite{alonso2008} and \cite{andrade2009}. We reproduce here one case:
the vibrational spectrum of an artificial benzene cristal (see
Fig.~\ref{fig:benzene-crystal}). Essentially, the calculations consist
of the time-propagation of the system, either with the standard CP
technique or with the modified Ehrenfest dynamics, for an interval of
time departing from a Boltzmann distribribution of velocities at a
given temperature. Then, the vibrational frequencies are obtained from
the Fourier transform of the velocity autocorrelation function.

Panels (b) and (c) display the serial and parallel computational cost,
respectively, of the two methods, defined as the computer time needed
to propagate one atomic unit of time. In the serial case, it can be
seen how for the system sizes studied, CP is more efficient; a
different scaling can already be guessed from the curve; indeed, if
these curves are extrapolated (inset), one can predict a crossing
point where the new Ehrenfest technique starts to be
advantageous. This is more patent in the parallel case, as can be seen
in panel (c). Panel (d) displays the different scalability of the two
methods: for a fixed system size, the system is equally divided in a
variable number of processors, and the figure displays the different
speed-ups obtained.

The key conclusion is that
the lack of the orthonormalization step permits a new efficient
parallelization layer, on top of the usual ones that are commonly
employed in CPMD (domain decomposition, and Brillouin zone K-points):
since the propagation step is independent for each orbital, it is
natural to parallelize the problem by distributing the KS states among
processors.  Communication is only required once per time-step to
calculate quantities that depend on a sum over states: the time
dependent densities and the forces over the ions.

\section{MD at finite electronic temperature}
\label{section:md-finite-temperature}

The previous section has addressed algorithmic alternatives to the solution of
the gsBOMD equations (\ref{eq:gsbomd-1}) and (\ref{eq:gsbomd-2}). These
represent the evolution of the classical nuclei, interacting all-to-all
through the potential $E_0(\vec{R})$. The resulting dynamics can be used to
calculate equilibrium averages at a given finite temperature, by assuming
ergodicity and computing time averages over a number of trajectories, once the
system has been appropriately coupled to a thermostat. The resulting marginal
equilibrium density in the nuclear positions space being, in the canonical
ensemble:
\begin{equation}
\label{eq:equilibriumBO}
p_{\rm gsBO}(\vec{R}) = \frac{\displaystyle {\rm e}^{-\beta E_0(\vec{R})}}
 {\int {\rm d}\vec{R}^\prime
  {\rm e}^{-\beta E_0(\vec{R}^\prime)}} \,,
\end{equation}
with $\beta:=1/k_B T$, or $\beta:=1/R T$ if per-mole units are used.

However, this scheme ignores completely the dynamics of the electrons, by
assuming that, even at a finite temperature, they are continuously tied to
their ground state. This assumption is legitimate if the electronic gap is
large compared to $k_B T$ at the temperature of interest. Indeed, in many
physical, chemical or biological processes the dynamical effects arising from
the presence of low lying electronic excited states have to be taken into
account. For instance, in situations where the Hydrogen bond is weak,
different states come close to each other and non-adiabatic proton transfer
transitions become rather likely at normal temperature~\cite{may2004}. In
these circumstances, the computation of ensemble averages cannot be based on a
model that assumes the nuclei moving on the ground-state BO surface.

In the DFT realm, the inclusion of electronic excited states in the dynamics
is very often done by working with partial occupation numbers to account for
the electronic
excitations~\cite{grumbach1994,alavi1994,alavi1995,marzari1997}, ideally
making use of temperature-dependent exchange and correlation
functionals~\cite{mermin1965,prodan2010,eschrig2010}. This scheme is however
tied to DFT, and is hindered by the difficulty of realistically approximating
this functional. Other alternative options are Ehrenfest dynamics and surface
hopping~\cite{tully1990} (for more on recent progress in non-adiabatic
electronic dynamics in mixed quantum-classical dynamics, see, for example,
\cite{zhu2005}).

Recently, Alonso et al.~\cite{alonso2010} have proposed a new alternative,
which can make use of the ability of TDDFT to compute electronic excited
states. In the following, we make a summary of the new technique.

In order to arrive to a general quantum-classical formalism, and to a suitable
expression for the quantum-classical equilibrium distribution that is
considered to be the correct one in the literature, it is preferable in this
case to follow the partial Wigner transformation route~\cite{wigner1932}, as
done by Ciccotti, Kapral and Nielsen~\cite{kapral1999,nielsen2001}. Let us
assume a quantum system of two particles of masses $m$ and $M$ ($M>m$) living
both in one dimension, whose canonical position and momentum operators are
$(\hat{x},\hat{p})$ and $(\hat{X},\hat{P})$, respectively. The generalization
to more particles and higher dimension is straightforward. Given an operator
$\hat{A}$, its partial Wigner transform $\hat{A}_W$ with respect to the
large-mass coordinate is defined as:
\begin{equation}
\hat{A}_W(X,P) := (2\pi\hbar)^{-1} \int\!\!{\rm d}z\; e^{\I Pz/\hbar} \langle X-z/2\vert \hat{A}\vert X+z/2\rangle\,.
\end{equation}

The operator $\hat{A}_W(X,P)$ acts on the Hilbert space of the \emph{light}
particle, and depends on the two real numbers $(X,P)$. It is possible to
reformulate all quantum theory in terms of these partial Wigner transforms; in
particular, if the Hamiltonian for the two particles is given by:
\begin{equation}
\hat{H} = \frac{\hat{P}^2}{2M} + \frac{\hat{p}^2}{2m} + V(\hat{x},\hat{X})\,,
\end{equation}
its transformation is:
\begin{equation}
\hat{H}_W(X,P) = \frac{P^2}{2M} + \frac{\hat{p}^2}{2m} + V(\hat{x},X)\,,
\end{equation}
i.e., one just has to substitute the quantum operators of the heavy
particle by the real numbers $(X,P)$.

If the state of the system is described by the density matrix $\hat{\rho}(t)$,
its evolution will be governed by von Neumann's equation,
\begin{equation}
\frac{\rm d}{{\rm d}t} \hat{\rho}(t) = -\frac{\I}{\hbar} \left[ \hat{H},\hat{\rho}(t)\right]\,,
\end{equation}
which can be cast into its partial Wigner-transformed form:
\begin{equation}
\label{eq:neumann-full}
\frac{\partial}{\partial t} \hat{\rho}_W = -\frac{\I}{\hbar} \left( \hat{H}_We^{\hbar\Lambda/2\I}\hat{\rho}_W -
\hat{\rho}_We^{\hbar\Lambda/2\I}\hat{H}_W \right)\,,
\end{equation}
where $\Lambda$ is the ``Poisson bracket operator'',
\begin{equation}
\Lambda := \overleftarrow{\frac{\partial}{\partial P}}\overrightarrow{\frac{\partial}{\partial X}} - 
\overleftarrow{\frac{\partial}{\partial X}}\overrightarrow{\frac{\partial}{\partial P}}\,,
\end{equation}
and the arrows indicate the direction in which each derivative acts.

Note that up to now, this is an exact reformulation of quantum mechanics (no
classical or semiclassical limit has been taken). However, this is also a
convenient departure point to take the classical limit for the heavy particle.
After an appropriate change of coordinates~\cite{kapral1999}, if we retain
only the first order terms in $\eta:=(m/M)^{1/2}$, Eq.~(\ref{eq:neumann-full})
is transformed into:
\begin{equation}
\label{eq:qcliouvillian}
\frac{\partial}{\partial t} {\hat{\rho}}_{\rm W} =
-\frac{i}{\hbar}\left[ \hat{H}_W,\hat{\rho}_W\right]
+ \frac{1}{2}\left( 
  \lbrace \hat{H}_W,\hat{\rho}_W \rbrace
- \lbrace \hat{\rho}_W, \hat{H}_W  \rbrace
\right)\,,
\end{equation}
where $\lbrace\cdot,\cdot\rbrace$ is the Poisson bracket with respect to the
canonical conjugate coordinates $(X,P)$,
\begin{equation}
\label{eq:Poissonbrackets}
\lbrace \hat{H}_W,\hat{\rho}_W \rbrace :=
 \frac{\partial \hat{H}_W}{\partial X}
 \frac{\partial \hat{\rho}_W}{\partial P}
-\frac{\partial \hat{H}_W}{\partial P}
 \frac{\partial \hat{\rho}_W}{\partial X}\,,
\end{equation}
and both $\hat{\rho}_W$ and $\hat{H}_W$ are functions of $(X,P)$.

The equilibrium density matrix in the partial Wigner representation at the
classical limit for the heavy particle, denoted by $\hat{\rho}_W^{eq}$ should
be stationary with respect to the evolution at first order in
$\eta:=(m/M)^{1/2}$ in Eq.~(\ref{eq:qcliouvillian}). If we use this property
and expand the equilibrium density matrix in powers of $\eta$:
\begin{equation}
\hat{\rho}_W^{\rm eq}(X,P) = \sum_{n=0}^{\infty} \eta^n
 \hat{\rho}_{W}^{\rm eq\;(n)}(X,P)\,,
\end{equation}
it can then be proved~\cite{nielsen2001} that the zero-th order term is given by:
\begin{eqnarray}
\label{eq:qcrho}
\hat{\rho}_{\rm W}^{\rm eq\;(0)}(X,P) =  \frac{1}{\mathcal{Z}} {\rm
  e}^{-\beta \hat{H}_W(X,P)}\,,
\end{eqnarray}
with
\begin{equation}
\mathcal{Z} := {\rm Tr}_{\rm q}\left[ \int {\rm d}X{\rm d}P {\rm
    e}^{-\beta\hat{H}_W(X,P)}\right]\,,
\end{equation}
the symbol ${\rm Tr}_{\rm q}$ meaning trace over the quantum degrees of freedom.

Note that~(\ref{eq:qcrho}) corresponds, at fixed classical variables $(X,P)$,
to the equilibrium density matrix \emph{for the electronic states}. However,
it is only an \emph{approximation} to the true quantum-classical equilibrium
density matrix, since it is not a stationary solution to the quantum-classical
Liouvillian given in Eq.~(\ref{eq:qcliouvillian}). This distribution is often
regarded, however, as the correct equilibrium distribution of the canonical
ensemble for a mixed quantum-classical system
\cite{mauri1993,parandekar2005,parandekar2006,schmidt2008,bastida2007}, and the average of
observables is computed as:
\begin{eqnarray}
\label{eq:averagesexact}
\langle \hat{O}(\hat{x},\hat{p},X,P)\rangle =  
{\rm Tr}_{\rm q}\int {\rm d}X{\rm d}P \hat{O}(\hat{x},\hat{p},X,P)
\hat{\rho}^{\rm eq \, (0)}_W(X,P) \,.
\end{eqnarray}

As mentioned, the careful analysis described in \cite{kapral1999,nielsen2001},
shows that this is a first order approximation in the square root of the
quantum-classical mass ratio $\eta:=(m/M)^{1/2}$, and therefore an acceptable
approximation if this ratio is small.

In the remaining part of this chapter, and following
\cite{alonso2010}, we will write a system of dynamic equations for the
classical particles such that the equilibrium distribution in the space of
classical variables is in fact given by Eq.~(\ref{eq:qcrho}). This is also a
goal of surface hopping methods \cite{tully1990}, although it is not fully
achieved since these methods do not exactly yield this distribution
\cite{schmidt2008}. We will
do this by deriving a temperature-dependent effective potential for the
classical variables, which differs from the ground-state potential energy
surface (PES) used in gsBOMD. It is straightforward, however, to write an
equation that gives the expression for the effective potential in terms of
this PES together with the BO PESs corresponding to the excited states of the
electronic Hamiltonian. Despite this property, it is worth remarking that the
approach described here is based on the assumption that the full system of
electrons and nuclei is in thermal equilibrium at a given temperature, and not
on the assumption that electrons immediately follow the nuclear motion (i.e.,
the adiabatic approximation), which is at the core of the BO scheme.

Let us assume that we are only interested in the average of observables that
depend explicitly only on the degrees of freedom of the heavy, classical
particle, $A=A(X,P)$. It is a matter of algebra (using Eqs.~(\ref{eq:qcrho})
and~(\ref{eq:averagesexact})) to prove that this average can be written as:
\begin{equation}
\label{averages}
\langle A(X,P)\rangle = \frac{1}{\mathcal{Z}}\int {\rm d}X{\rm d}P A(X,P) {\rm e}^{-\beta H_{\rm eff}(X,P;\beta)}\,,
\end{equation}
where we have introduced an \emph{effective} Hamiltonian $H_{\rm eff}$,
defined as:
\begin{equation}
\label{Heff-mix}
H_{\rm eff}(X,P;\beta) := -\frac{1}{\beta}
\ln{ {\rm Tr}_{\rm q}{\rm e}^{-\beta H_W(X,P)} }\,.
\end{equation}

The partition function $\mathcal{Z}$ can also be written in terms of
the effective Hamiltonian:
\begin{equation}
\mathcal{Z} = \int {\rm d}X{\rm d}P {\rm e}^{-\beta H_{\rm eff}(X,P;\beta)}\,,
\end{equation}

Hence, the quantum subsystem has been ``integrated out'', and does not appear
explicitly in the equations any more (of course, it has not disappeared, being
hidden in the definition of the effective Hamiltonian). In this way, the more
complicated quantum-classical calculations have been reduced to a simpler
classical dynamics with an appropriate effective Hamiltonian, which produces
the same equilibrium averages of classical observables [Eq.~(\ref{averages})]
as the one we would obtain using Eq.~(\ref{eq:qcrho})
in~(\ref{eq:averagesexact}), and hence incorporates the quantum back-reaction
on the evolution of the classical variables, at least at the level of
equilibrium properties.

In the case of a molecular system, the total (partially Wigner transformed)
Hamiltonian reads:
\begin{equation}
\hat{H}(\vec{R},\vec{P}) =
 T_{\rm n}(\vec{P}) + \hat{H}_{\rm e}(\vec{R})\,,
\end{equation}
where $\vec{R}$ denotes collectively all nuclear coordinates, $\vec{P}$ all
nuclear momenta, $T_n(\vec{P})$ is the total nuclear kinetic energy, and
$\hat{H}_{\rm e}(\vec{R})$ is the electronic Hamiltonian in Eq.~(\ref{eq:He}),
that includes the electronic kinetic term and all the interactions. The
effective Hamiltonian, defined in Eq.~(\ref{Heff-mix}) in general, is in this
case of a molecular system given by:
\begin{equation}
\label{eq:Heff}
H_{\rm eff}(\vec{R},\vec{P};\beta) := T_{\rm n}(\vec{P}) - 
 \frac{1}{\beta}\ln{ {\rm Tr}_{\rm q}
 {\rm e}^{-\beta \hat{H}_{\rm e}(\vec{R})} }
 =: T_{\rm n}(\vec{P}) + V_{\rm eff}(\vec{R};\beta)\,,
\end{equation}
where the last equality is a definition for the \emph{effective}
potential $V_{\rm eff}(\vec{R};\beta)$.

Now, making use of the adiabatic basis, defined in
Eq.~(\ref{eq:adiabatic-basis}) as the set of all eigenvectors of electronic
Hamiltonian $\hat{H}_{\rm e}(\vec{R})$, we can rewrite $V_{\rm
eff}(\vec{R};\beta)$ as:
\begin{equation}
\label{eq:veff}
V_{\rm eff}(\vec{R};\beta) = 
 E_0(\vec{R}) - \frac{1}{\beta}\ln{
  \left[1+\sum_{n>0}{\rm e}^{-\beta E_{n0}(\vec{R})}\right]}\,,
\end{equation}
where $E_{\rm n0}(\vec{R}) := E_n(\vec{R})-E_0(\vec{R})$. It is for
the computation of these excitation energies that TDDFT can be
employed. The proposed dynamics would be, therefore, \emph{based} on
TDDFT. Of course, any other many-electron technique can also be used.

This equation permits to see explicitly how the ground state energy $E_0$
differs from $V_{\rm eff}$, and in consequence how a MD based on $V_{\rm eff}$
is going to differ from a gsBOMD. In particular, notice that $V_{\rm
eff}(\vec{R};\beta) \le E_0(\vec{R})$, and compare the marginal probability
density in the gsBOMD case in Eq.~(\ref{eq:equilibriumBO}) to the one produced
using the new dynamics:
\begin{equation}
\label{eq:equilibriumeff}
p_{\rm eff}(\vec{R}) = \frac{\displaystyle 
 \left( 1+\sum_{n>0}{\rm e}^{-\beta E_{n0}(\vec{R})} \right)
 {\rm e}^{-\beta E_0(\vec{R})}}
 {\displaystyle \int {\rm d}\vec{R}^\prime
  \left( 1+\sum_{n>0}{\rm e}^{-\beta E_{n0}(\vec{R^\prime})} \right)
  {\rm e}^{-\beta E_0(\vec{R}^\prime)}} \,.
\end{equation}

Finally, note that to the extent that nuclei do not have quantum behavior near
conical intersections or spin crossings, nothing prevents us to use this
equation also in these cases.

The definition of the classical, effective Hamiltonian for the nuclear
coordinates in Eq.~(\ref{eq:Heff}) allows us now to use any of the
well-established techniques available for computing canonical equilibrium
averages in a classical system. Of course, since $H_{\rm eff}$ in
Eq.~(\ref{eq:Heff}) depends on $T$, any Monte Carlo or dynamical method must
be performed at the same $T$ that $H_{\rm eff}$ was computed in order to
produce consistent results, given in this case by the convenient
expression~(\ref{averages}). For example, we could use (classical) Monte Carlo
methods, or, if we want to perform MD simulations, we could propagate the
stochastic Langevin dynamics associated to the Hamiltonian (\ref{eq:Heff}):
\begin{equation}
\label{langevin-eq}
M_J\ddot{\vec{R}}_J(t) = -\vec{\nabla}_J V_{\rm eff}(\vec{R}(t);\beta) - M_J\gamma\dot{\vec{R}}_J(t) + M_J\vec{\Xi}(t)\,,
\end{equation}
where $\vec{\Xi}$ is a vector of stochastic fluctuations, obeying $\langle
\Xi_i(t)\rangle = 0 $ and $\langle \Xi_i(t_1)\Xi_j(t_2)\rangle = 2\gamma k_B T
\delta_{ij}\delta (t_1-t_2)$ which relates the dissipation strength $\gamma$
and the temperature $T$ to the fluctuations (fluctuation-dissipation theorem).

Indeed, it is well-known that this Langevin dynamics is equivalent to the
Fokker-Planck equation for the probability density $W(\vec{R},\vec{P})$ in the
classical phase space \cite{vankampen2007}:
\begin{align}
\nonumber
\label{FP-gral}
 \frac{\partial W (\vec{R},\vec{P};t)}{\partial t} & = 
\{H_{\rm eff}(\vec{R},\vec{P};\beta), W (\vec{R},\vec{P};t) \} 
\\ & + \gamma \sum_J 
\partial_{\vec {P}_J} (\vec {P}_J  + M k_{\rm B} T \partial_{\vec {P}_J} )  
W(\vec{R},\vec{P};t)\,.
\end{align}

Any solution to Eq.~(\ref{FP-gral}) approaches at infinite time a distribution
$W_{\rm eq}(\vec{R},\vec{P})$ such that $\partial_t W_{\rm
eq}(\vec{R},\vec{P}) = 0$. This stationary solution is unique and equal to the
Gibbs distribution, $W_{\rm eq}(\vec{R},\vec{P}) = \mathcal{Z}^{-1} \; {\rm
e}^{-\beta H_{\rm eff}(\vec{R},\vec{P};\beta)}$ \cite{vankampen2007}. Thus,
the long-time solutions of Eq.~(\ref{FP-gral}), and hence those of
Eq.~(\ref{langevin-eq}) reproduce the canonical averages in
Eq.~(\ref{averages}). This property, which is also satisfied by other dynamics
like the one proposed by Nos\'e~\cite{nose1984,nose2001} if the $H_{\rm eff}$
in Eq.~(\ref{eq:Heff}) is used, comes out in a very natural way from the
present formalism while it is yet unclear of other ab initio MD candidates for
going beyond gsBOMD \cite{mauri1993,parandekar2005,schmidt2008,bastida2007}.

When would this new MD scheme be useful? The approach introduced in this
section is particularly suited to the case of conical intersection or
spin-crossing \cite{yarkony1996}, since it does not assume that the electrons
or quantum variables immediately follow the nuclear motion, in contrast to any
adiabatic approach. Another interesting application pertains the debated issue
of quantum effects in proton transfer~\cite{iyengar2008}. It is a matter of
current debate to what extent protons behave ``quantum-like'' in biomolecular
systems (e.g. is there any trace of superposition, tunneling or entanglement
in their behavior?). Recently, McKenzie and coworkers~\cite{bothma2010} have
carefully examined the issue, and concluded that ``tunneling well below the
barrier only occurs for temperatures less than a temperature $T_0$ which is
determined by the curvature of the PES at the top of the barrier.'' In
consequence, the correct determination of this curvature is of paramount
importance.

The curvature predicted by the temperature-dependent effective potential
introduced here is smaller than the one corresponding to the ground state PES,
in the cases in which the quantum excited surfaces approach, at the barrier
top, the ground state one. Therefore, $T_0$ would be smaller than that
corresponding to the ground state PES (see Eq. (8) in~\cite{bothma2010}), and
hence the conclusion in this reference ``that quantum tunneling does not play
a significant role in hydrogen transfer in enzymes'' is reinforced by the
results of the new dynamics.

\section*{Acknowledgments}
J. L. Alonso, A. Castro and P. Echenique aknowledge support from the
research grants E24/3 (DGA, Spain), FIS2009-13364-C02-01 (MICINN,
Spain). P. Echenique aknowledges support from the research grant
200980I064 (CSIC, Spain). A. Rubio acknowledges funding by the Spanish
MEC (FIS2007-65702-C02-01), ACI-promciona project (ACI2009-1036),
``Grupos Consolidados UPV/EHU del Gobierno Vasco'' (IT-319-07), the
European Research Council through the advance grant DYNamo (267374),
and the European Community through projects e-I3 ETSF (Contract
No. 211956) and THEMA (228539).


\end{document}